# Helium Abundance Periods Observed by the Solar Probe Cup on Parker Solar Probe: Encounters 1–14

Madisen Johnson[1,2,3], Yeimy J. Rivera[1], Tatiana Niembro[1], Kristoff Paulson[1], Samuel T. Badman[1], Michael L. Stevens[1], Isabella Dieguez[1,4], Anthony Case[1,5], Stuart D. Bale[6,7], and Justin Kasper[5,8]

[1] Center for Astrophysics | Harvard & Smithsonian, 60 Garden Street, Cambridge, MA 02138, USA
[2] Department of Physics and Astronomy, Rutgers University, Piscataway, NJ 08854-8019, USA
[3] Department of Astronomy, University of Wisconsin-Madison, Madison, WI 53706-1507, USA
[4] Department of Physics, University of Miami, Miami, FL 33146, USA
[5] BWX Technologies Inc., Washington, DC 20001, USA
[6] Physics Department, University of California, Berkeley, CA 94720-7300, USA
[7] Space Sciences Laboratory, University of California, Berkeley, CA 94720-7450, USA
[8] University of Michigan, Ann Arbor, MI 48109, USA

Received 2023 December 7; revised 2024 January 26; accepted 2024 January 30; published 2024 March 19

## Abstract

Parker Solar Probe is a mission designed to explore the properties of the solar wind closer than ever before. Detailed particle observations from the Solar Probe Cup (SPC) have primarily focused on examining the proton population in the solar wind. However, several periods throughout the Parker mission have indicated that SPC has observed a pronounced and distinctive population of fully ionized helium, $He^{2+}$. Minor ions are imprinted with properties of the solar wind's source region, as well as mechanisms active during outflow, making them sensitive markers of its origin and formation at the Sun. Through a detailed analysis of the $He^{2+}$ velocity distributions functions, this work examines periods where significant and persistent $He^{2+}$ peaks are observed with SPC. We compute the helium abundance and examine the stream's bulk speed, density, temperature, magnetic field topology, and electron strahl properties to identify distinctive solar-wind features that can provide insight to their solar source. We find that nearly all periods exhibit an elevated mean helium composition (8.34%) compared to typical solar wind and a majority (∼87%) of these periods are connected to coronal mass ejections (CMEs), with the highest abundance reaching 23.1%. The helium abundance and number of events increases as the solar cycle approaches maximum, with a weak dependence on speed. Additionally, the events not associated with a CME are clustered near the heliospheric current sheet, suggesting they are connected to streamer belt outflows. However, there are currently no theoretical explanations that fully describe the range of depleted and elevated helium abundances observed.

*Unified Astronomy Thesaurus concepts:* Solar abundances (1474); Solar coronal mass ejections (310); Solar wind (1534); Solar physics (1476)

## 1. Introduction

The solar wind is formed through a continuous outflow of plasma escaping the solar corona that engulfs interplanetary space. The properties of the solar wind and transients measured in the heliosphere have been shown to be quite variable in velocity, temperature, and density as well as in its ion and elemental composition (Geiss et al. 1995; McComas et al. 2000; von Steiger et al. 2000). The ion composition measured in the heliosphere is a reflection of the plasma's thermodynamic evolution across the corona, and the charge-state composition can provide a link to the plasma's coronal origin (Rakowski et al. 2007; Gruesbeck et al. 2011; Landi et al. 2012a; Rivera et al. 2019). The distinct elemental composition observed across different solar-wind streams and ejecta is a product of fractionation processes at their source region such as the first ionization potential (FIP) effect (Laming 2015) and gravitational settling (Weberg et al. 2012; Rivera et al. 2021). From decades of coronal observations, we observe the FIP effect to different degrees across coronal structures (Pottasch 1963). The FIP effect results in the overall enhancement, of a factor 3 or above, of low-FIP elements in the corona (below a 10 eV threshold) relative to the Sun's photospheric composition, while the elements with a FIP value above this limit remain relatively unaffected (Feldman & Laming 2000). Another important fractionation effect, observed both remotely and in situ, is gravitational settling, which depletes coronal plasma composition in accordance to particle mass. The various heavy-ion features that remain imprinted in the plasma once it leaves the corona provide insight to processes active in the corona. Therefore, the ion and elemental composition measured throughout the heliosphere are important tracers of the eruption processes (Lynch et al. 2011; Rakowski et al. 2011; Laming et al. 2023; Rivera et al. 2023), solar-wind release and dynamic outflow (Landi et al. 2012b; Scott et al. 2022), and properties of the plasma's source region (Xu & Borovsky 2015; Zurbuchen et al. 2016; Ervin et al. 2023; Lynch et al. 2023).

In particular, coronal mass ejections (CMEs) have been shown to contain some of the most extreme ionization stages as well as the largest differences in their compositional makeup as compared to the solar wind and Sun's photosphere (Zurbuchen & Richardson 2006; Zurbuchen et al. 2016), with some of the most extreme examples being the Bastille Day and Halloween events of solar cycle 23 (Liu et al. 2008; Rivera et al. 2023). This is likely due to their explosive release and rapid acceleration from the Sun that evacuates significant amounts of hot, dense active-region material and which are known to contain the most extreme low-FIP enhancements and the hottest coronal plasma (Feldman & Laming 2000; Widing & Feldman 2001). In addition, CMEs are also observed to be significantly heated during the early stages of propagation (Landi et al. 2010). As such, CMEs are measured in the heliosphere to contain highly ionized ions, where iron ions of charge state $\geqslant 16+$ are routinely observed during their passage, and can be successfully used to distinguish their appearance among the continuous flow of solar wind, which is less ionized (Bame et al. 1979; Fenimore 1980; Ipavich et al. 1986; Lepri et al. 2001; Lepri & Zurbuchen 2004). CMEs reflect elevated average iron charge-state values, $\langle Q_{\text{Fe}} \rangle > 10+$, that are much higher than typical solar wind (Rivera et al. 2022).

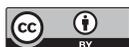

Similarly, their elemental composition can exhibit some of the highest changes from photospheric values. Statistical properties near the Earth (1 au) show that CMEs have on average a helium abundance, defined as $A_{\text{He}} = N_{\text{He}}/N_{\text{H}} \times 100$, greater than 8%, where $N_{\text{He}}$ and $N_{\text{H}}$ is the helium and hydrogen number density (e.g., Hirshberg et al. 1972; Borrini et al. 1982). For reference, the photospheric abundances reported are $A_{\text{He}} \sim 8.4\%$ (Asplund et al. 2021). In fact, CMEs exhibit the highest helium abundances, reaching up to $A_{\text{He}} = 15\%-20\%$, which is significantly enhanced compared to the solar photospheric composition (Song et al. 2022). Enhanced helium abundances can also be observed at flare sites during their decay phase, reaching $A_{\text{He}} = 12.2\% \pm 2.4\%$ (Feldman et al. 2005). These $A_{\text{He}}$ values are in large contrast to the solar wind, which is primarily depleted compared to photospheric values (Ogilvie & Hirshberg 1974; Aellig et al. 2001; Kasper et al. 2007, 2012; Alterman & Kasper 2019; Alterman et al. 2021; Yogesh & Srivastava 2021). Moreover, recent studies find that, like the solar wind, CMEs also exhibit a solar-cycle dependence in their composition (Song et al. 2021, 2022; Li et al. 2023). The changes in abundance have been connected to differences among their source regions (i.e., active regions in comparison to quiet Sun) whose thermal structure, elemental composition, and appearance across the solar cycle can be markedly different (see recent summary in Norton et al. 2023). Differences in elemental composition can additionally be attributed to the appearance of gravitational settling and ion dropouts, which can fractionate the plasma; both these effects also exhibit a solar-cycle dependence (Weberg et al. 2012; Rivera et al. 2022; Zhao et al. 2022).

Given that helium has the highest FIP value, 24.6 eV, its enhancement is likely produced by a process, or processes, different than the FIP effect. Having a FIP value well above 10 eV, helium is a high-FIP element that would be expected to exhibit abundance levels similar to that in the photosphere. Yet its abundance can vary significantly (Kasper et al. 2007; Song et al. 2022). The variability of helium abundance observed in situ in the solar wind and transients is directly connected to processes established at its source region or during an eruption. However, we do not have a good model to explain the process (es) which cause(s) helium abundances to vary to such a high degree. Previous work has connected the changes in helium abundance to source surface footpoint magnetic field magnitude and Alfvén speed (Wang 2008), and theorized to come about through variation in wave–particle interactions (Kasper et al. 2007), Coulomb friction (Geiss et al. 1970), or gravitational stratification (Laming & Feldman 2001). To better understand the significant range of helium abundances, between <1% to exceeding 20% across the solar wind and transients observed throughout the heliosphere, it is necessary to examine the solar wind and transient structures closer to the Sun, where plasma has had less time to evolve, and remains more closely tied to the signatures of their coronal sources.

To gain insight into the helium abundances observed in the inner heliosphere, at some of the closest distances sampled to date, we can assess solar wind observed with the Parker Solar Probe (hereafter Parker). In this work, we systematically examine the $\text{He}^{2+}$ ion population that is intermittently measured with the Solar Probe Cup (SPC) on board Parker. Measurements are taken throughout the inner heliosphere encompassing periods in and out of encounters 1–14 to comprehensively characterize periods with a pronounced and distinctive $\text{He}^{2+}$, or alpha-particle, population. We present a list of periods where helium abundances can be accurately determined by simultaneously fitting Maxwellian curves to the proton and alpha population measured by SPC. We also report on the properties of the stream in those periods, such as its connection to transients or ambient solar wind, the statistical properties of the helium abundances across the solar cycle, so far, and dependence on bulk speed. We find helium abundances determined with SPC are generally enriched compared to established trends at 1 au. The high helium abundances reflected by the list of periods suggests that SPC is the most sensitive to plasma with the highest helium content, such as is routinely observed in CMEs. This makes the high-resolution SPC measurements ideal for examining the fine character of newly ejected CME substructure, i.e., fragmented prominence features and neighboring hot, active-region outflow, early in its evolution, which can be better connected to features captured in extreme-ultraviolet and coronagraph images.

The paper is organized as follows. Section 2 discusses the instruments and data used for the analysis. Section 3 discusses the fitting routine and helium abundance calculation. Section 4 presents the catalog of periods identified with some statistical results. Lastly, Section 5 summarizes the main conclusions.

## 2. Observations

To characterize the helium abundances as well as provide context to the solar-wind stream in which they are embedded, this study utilizes measurements from the Solar Wind Electrons Alphas and Protons (SWEAP; Kasper et al. 2016) instrument, which consists of the SPC (Case et al. 2020) and the Solar Probe ANalyzers (SPAN; Whittlesey et al. 2020). We also utilize the FIELDS (Bale et al. 2016) suite of instruments aboard Parker (Fox et al. 2016). Using these instruments, we analyzed periods of time in encounters 1–14 from 2018 to the end of 2022.

To compute the helium composition of the solar wind, we use measurements from the SPC, which provides velocity distribution functions (VDFs) of the thermal component of the solar wind. Unlike all other plasma instruments on board, SPC points directly at the Sun, capturing the outflowing solar wind without the field-of-view constraint seen by SPAN due to the position of the instrument on the side of Parker. The VDFs are dominated by the presence of protons; however, they can occasionally contain clear signatures of heavy ions at higher energies, the most abundant and unambiguous being $\text{He}^{2+}$ (also known as alpha particles). During these periods, the VDFs can then be fit with one Maxwellian curve for the proton population and one Maxwellian curve for the alpha population. These fits can then be used to determine proton and alpha velocity, density, and temperature where the $He^{2+}$ appears visible above the noise. The helium abundance, defined as $A_{He} = N_{He2+}/N_p \times 100$, is computed as the ratio of alpha-particle density to the proton density. Details of the procedure are presented in Section 3.

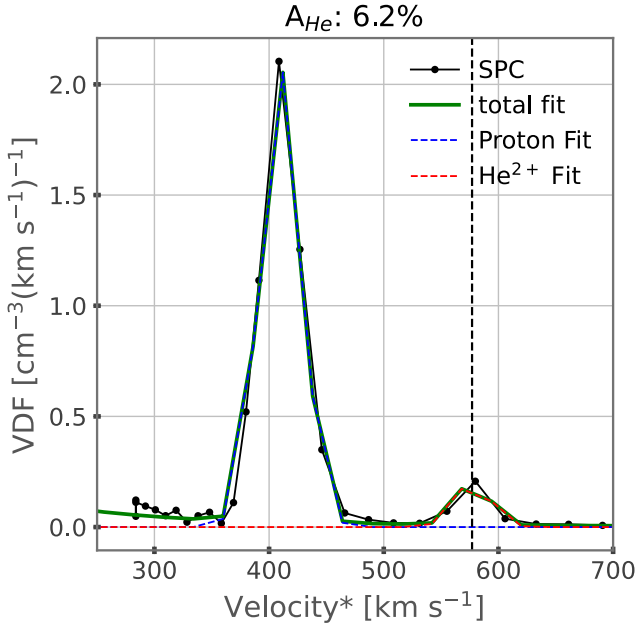

**Figure 1.** Example of an SPC VDF profile with the two Maxwellian curve fits used to compute the properties of the proton and helium populations.

To analyze the magnetic fields of periods where helium abundances are computed, we use observations from FIELDS. The FIELDS suite provides 3D magnetic field components, measured by the fluxgate magnetometer that captures the local magnetic field topology within the solar wind. For our analysis, we use the data collected at four vectors cycle$^{-1}$. SPAN-e provides pitch-angle distributions of electrons, which can help further characterize magnetic field structure and connection to the Sun. For instance, electron bidirectionality is a clear signature of a field line that is anchored to the Sun at both footpoints, often observed during heliospheric current sheet (HCS) crossings or transient events, such as CMEs (Richardson & Cane 2010; Owens & Forsyth 2013).

For the purposes of our study, we assume that the flow angle of the protons is the same as that of the alpha particles. This is due to the relatively low signal-to-noise ratio of the helium peak, which does not allow us to estimate the flow angle of the $He^{2+}$ more accurately than when assuming that it is the same as the proton flow angle. Additionally, when the temperature of the two particles is equal equilibrium is implied, which strengthens the assumption that the flow angles would be the same for the proton and alpha particles.

## 3. $He^{2+}$ Fitting Analysis

To examine the specific solar-wind structures associated with elevated helium content, we compute the helium abundance and analyze the stream's associated bulk speed, temperature, and its magnetic field topology and electron pitch-angle distribution properties. We use the properties observed to determine the helium content of the associated transient or ambient solar wind to compare with statistical values observed at 1 au.

To analyze the helium content in the solar wind, we examine the VDFs measured by SPC, where distinctive $He^{2+}$ distributions are observed. A VDF is generated from the differential energy flux determined by the total current on the four collector plates on SPC, which is combined and divided by the effective area. For SPC, the VDFs are only measured along the flow direction into the instrument. This 1D VDF is known as a reduced distribution function (RDF). Following Case et al. (2020), the RDF is calculated as approximately $D/q\bar{v}\delta v$, where $D$ is the differential charge flux density, $q$ is the charge, $\bar{v}$ is the average equivalent velocity of an ion during a voltage window measurement, and $\delta v$ is the effective width of a voltage window in velocity units. The differential charge flux density, $D$, in charge per second is measured across a 1 cm$^2$ unit area for ions with kinetic energy between the voltage window, $V_{hi} < mv^2/2q < V_{lo}$, where $Vq = 1/2 mv^2$.

Figure 1 shows an example of an RDF, or equivalently a 1D VDF, on 2021 April 25 that contains a proton population whose distribution peaks at $v_p \simeq 400$ km s$^{-1}$ and a $He^{2+}$ population peaking at a speed close to $v_p^*\sqrt{2}$, as expected for comoving particles with a charge-to-mass ratio of 1/2. Since the SPC detector assumes all particles hitting the instrument are protons, the figure denotes the velocity as "Velocity*," indicating a velocity computed as velocity* $= \sqrt{2Vq_e/(m_p)}$, where $V$ is the voltage, $q_e$ is the electron charge, and $m_p$ is the proton mass. The $He^{2+}$ population will be $m_a = 4 \times m_p$, and $q_e$ is 2. Therefore, the peak in the alpha population is seen at the velocity of the proton speed divided by $\sqrt{2}$ in the VDFs ($v_a$ = velocity* $\times \sqrt{2}$) while $v_p$ = velocity*, as discussed in Case et al. (2020). This velocity threshold also serves as means of distinguishing the $He^{2+}$ from a field-aligned proton beam population, which can appear as an additional peak when the interplanetary magnetic field is strongly radial.

To examine each population, we implement two Maxwellian functions to fit the individual proton and alpha-particle VDFs, defined as

$$f(v) = \frac{\rho_p}{\sqrt{\pi}\,v_{th}} e^{\frac{-(v-v_p)^2}{v_{th}^2}} + \frac{\rho_a}{\sqrt{\pi}\,v_{th}} e^{\frac{-(v-v_a)^2}{v_{th}^2}}, \quad (1)$$

where $\rho$ is the integral over the fitted peak, $v_{th}$ is the thermal velocity as defined by the width of the peak being fitted, $v_a$ is the alpha-particle velocity, and $v_p$ is the proton velocity.

Additionally, we also implemented a fitting parameter as a zeroth-order correction for the background noise in the observations. The fit is $f(v)^2 + \text{noise}^2$, where the noise is of the form of a VDF as $s/(v \times \delta v)$, where $\delta v$ is the effective width of a voltage window in velocity units, and $s$ is a parameter for the noise in charge flux units.

Using the fit of these two Maxwellian curves with noise capturing the proton and the alpha particles, we compute the density, temperature, and velocity of each population. Using the individual fit properties, we can compute the helium abundances by taking the ratio of the alpha to proton number density.

In order to ensure reasonable Maxwellian fits, we employed the total chi-squared metric to quantify the goodness of fit:

$$\chi^2 = \sum \frac{(O_i - E_i)^2}{E_i}, \quad (2)$$

where $O$ is the observed values, $E$ is the expected values, and the index $i$ represents the individual points used for the calculation. We divide our chi-squared by our degrees of freedom (here two), calculated as the number of independent points (five), minus the number of fit parameters (three). We compute the chi-squared value for two points to the left and right of both the proton and alpha distribution peaks. Fits with a total chi-squared value greater the threshold ($\chi^2 \geqslant 1$), are deemed to be poor fits to the data and are not included in the analysis. Each period included in the analysis is also examined via visual inspection.

We note that the fits will favor higher-density, higher-velocity, and relatively lower-temperature periods, which we discuss in the Results (Section 4). The individual proton and alpha-particle peaks are more identifiable where there is an obvious minimum between the proton and alpha-particle VDFs. As such, periods of a high field-aligned proton beam population with a strong radial field will also generate some ambiguity. This suggests that relatively low-alpha-density periods will be more difficult to identify cleanly as the "good" fits favor higher alpha densities that rise well above above the noise floor, higher bulk speeds allowing the distributions of the two population peaks to be further separated and more distinguishable, and lower temperatures where the distributions exhibit a narrower spread and high peak with less potential overlap between the two populations. We also note that because there are several factors which contribute to identifying and fitting the alpha population, i.e., particle flux, temperature, speed, as well as the noise level, there is no exact minimum threshold for the abundances that can measured. Instead, it is based on the conditions mentioned.

## 4. Results

After establishing the algorithm, we implemented the scheme across the Parker data set, that is, between encounters 1–14, including data inside and outside of encounter periods. All events identified through this method are listed in Table 1. The table includes a categorization between ambient wind (SW) and CMEs/CME*, as is discussed later in the text. If the period is associated with a CME/CME*, we also include an eruption date and time at the Sun. We list the time frame of the helium abundances measured at Parker in "Observed at Parker," the mean helium abundance as $A_{\mathrm{He}}$, the mean bulk proton speed in kilometers per second, and the mean heliocentric distance of Parker during the observation. We include whether the periods were associated with a notable rotation of the magnetic field in connection to a helical structure, as well as if any bidirectional electron signatures were present at any point during the helium abundance signature. We also list if the CME was identified in the Helio4Cast living catalog (Möstl et al. 2017, 2020) or simulated by the Community Coordinated Modeling Center (CCMC) via the Space Weather Database Of Notifications, Knowledge, Information (DONKI) and observed within 12 hr of its forecasted arrival. Lastly, we include notes relating to published event studies targeting these events or periods and potential HCS crossings nearby. Therefore, if the event appears within the boundaries of a CME as identified by the Helio4Cast or in connection to a DONKI simulation, its category is labeled as a CME, while if the period shows the principal character of a CME, bidirectional electrons and a helical structure, but is not on the Helio4Cast or DONKI simulated list of events, it is labeled as a CME*. The CME* category strongly suggests that the helium signature observed at Parker is likely to be an unidentified CME.

Through this analysis, we found that a majority, 87%, of periods where the $\mathrm{He}^{2+}$ population and peak was clearly distinguishable in SPC occurred during a CME/CME*. The CMEs were identified through a collection of signatures: the character of the local magnetic field, such as bidirectional electrons and helical magnetic field structures, as well as obvious discontinuities in the bulk properties of the plasma such as sharp changes in density, temperature, and velocity (Richardson & Cane 2010). CME identification was corroborated with the Helio4Cast CME list observed by Parker as well as the DONKI simulation database.

Periods of solar-wind classification are nontransient events showing no obvious CME signatures. The number of these events are small, only 13% of the total number of periods listed. As discussed, SPC and the associated fitting routine will be sensitive to plasma with the largest helium content where the alpha population is clearly distinguishable against the background and dominating proton population. Therefore, we expect the largest contribution of the events identified to be helium-rich plasma, such as CMEs.

Besides simply identifying intervals of elevated helium abundances, SPC measurements allow a detailed analysis of the substructure of these enhancements and associated solar-wind streams in unprecedented resolution and closer to the Sun than ever before. In the following sections, we examine this substructure for an example CME, a HCS crossing, and a solar-wind (nontransient) enhancement event.

### 4.1. Coronal Mass Ejections

Figure 2 shows a confirmed CME observed on 2021 November 11, as listed in the Helio4Cast database and identified through the Space Weather simulations from DONKI. The figure, from top to bottom, shows the SPC particle flux, helium-to-proton ratio as $N_{\mathrm{He}}/N_{\mathrm{H}}$, bulk speed, density, thermal speed, magnetic field components and magnitude, and pitch-angle distribution of the electron flux at 314.45 eV. The CME boundary is between 2021 November 9 18:40 UT and 2021 November 10 04:38 UT as identified by Helio4Cast. At the front, the plasma shows a steady <5% helium abundance in the main sheath region as characterized by the large variability in the magnetic field components. The sheath is generally associated with compressed solar-wind material at the CME front, in line with that helium content (Zurbuchen & Richardson 2006; Richardson & Cane 2010). During this time, we also see an increase in the velocity of the protons as density and thermal speed increases. However, no significant proton density enhancement is observed.

The helium abundance is gradually enhanced after the sheath, with a mean value of 7.34% and values ranging from 2% and 15%. These values are higher than what we expect from the ambient solar wind and align with previous helium measurements in CMEs showing >8% (Zurbuchen & Richardson 2006). The elevated helium abundances appear connected to the flux rope. After the sheath, the magnetic field becomes smooth, indicating significant expansion, while some rotation in the field and enhancement in the magnetic field magnitude is observed, suggesting flux-rope structure.

The Astrophysical Journal, 964:81 (12pp), 2024 March 20Johnson et al.

5Table 1
He$^{2+}$ Periods and Their Properties Observed by Parker Solar Probe

| Category | Eruption Date (YYYYMMDD 00:00UT) | Observed at Parker (YYYYMMDD 00:00UT) | $A_{\mathrm{He}}{}^{+\sigma}_{-\sigma}$ (%) | $v_{\mathrm{bulk}}$ (km s$^{-1}$) | Heliocentric Distance ($R_\odot$) | Helical Structure | Bidirectional Electrons | Helio4Cast | CCMC | Notes and References |
|---|---|---|---|---|---|---|---|---|---|---|
| CME | 20181110 20:00 | 20181113 07:10–23:59 | $3.21^{4.52}_{1.91}$ | 286.6 | 62.0 | x | x | x | x | Giacalone et al. (2020); Korreck et al. (2020); Nieves-Chinchilla et al. (2020); Good et al. (2023) Listed SEP event: Mitchell et al. (2023) |
| CME | ⋯ | 20190324 03:30–23:59 | $3.80^{6.05}_{1.57}$ | 360.0 | 81.4 | x | ⋯ | x | ⋯ | Good et al. (2023) Listed SEP event: Mitchell et al. (2023) |
| SW | ⋯ | 20190327 19:10–23:59 | $1.93^{2.52}_{1.36}$ | 358.9 | 67.3 | x | ⋯ | ⋯ | ⋯ | Near HCS crossing Listed SEP event: Mitchell et al. (2023) |
| CME | 20200528 01:25 | 20200528 09:00–20:00 | $2.57^{3.82}_{1.33}$ | 323.0 | 75.6 | x | x | x | x | Möstl et al. (2022); Good et al. (2023); Cheng et al. (2023) Listed SEP event: Mitchell et al. (2023) |
| SW | ⋯ | 20200613 09:10–23:59 | $3.37^{5.66}_{1.08}$ | 286.5 | 55.3 | ⋯ | x | ⋯ | ⋯ | Listed SEP event: Mitchell et al. (2023) |
| CME* | ⋯ | 20200921 00:00–23:59 | $2.98^{4.00}_{1.97}$ | 313.4 | 55.1 | x | x | ⋯ | ⋯ | ⋯ |
| SW | ⋯ | 20201226 00:00–23:59 | $4.65^{6.08}_{3.24}$ | 311.7 | 129.1 | x | ⋯ | ⋯ | ⋯ | Near HCS crossing |
| CME | 20210327 13:40 | 20210330 00:00–23:59 | $7.36^{10.5}_{4.20}$ | 390.0 | 148.0 | x | x | x | ⋯ | ⋯ |
| CME | ⋯ | 20210425 00:30–12:30 | $4.89^{6.63}_{3.14}$ | 338.0 | 42.8 | x | ⋯ | ⋯ | ⋯ | Niembro et al. (2023) |
| CME | 20210725 13:36 | 20210725 23:00–20210726 23:59 | $5.64^{7.21}_{4.07}$ | 417.2 | 103.9 | x | x | ⋯ | x | Listed SEP event: Mitchell et al. (2023) |
| CME | 20210728 20:24 | 20210730 17:50–20210801 23:59 | $7.09^{9.44}_{4.72}$ | 449.2 | 82.8 | x | x | ⋯ | x | Listed SEP event: Mitchell et al. (2023) |
| CME | 20210803 16:04 | 20210804 11:50–20:30 | $4.40^{7.00}_{1.80}$ | 315.0 | 52.3 | x | x | ⋯ | x | ⋯ |
| CME | ⋯ | 20210912 16:00–23:59 | $6.97^{8.82}_{5.12}$ | 347.6 | 154.7 | x | ⋯ | x | ⋯ | ⋯ |
| CME | 20210915 19:47 | 20210916 17:00–23:59 | $4.30^{5.69}_{2.91}$ | 312.6 | 160.3 | x | x | ⋯ | x | ⋯ |
| CME | 20210923 20:37 | 20210927 03:00–23:59 | $9.76^{12.7}_{7.27}$ | 345.4 | 168.0 | x | x | x | x | ⋯ |
| CME* | ⋯ | 20211013 02:00–14:00 | $10.3^{12.7}_{7.76}$ | 403.2 | 160.4 | x | x | ⋯ | ⋯ | Good et al. (2023) |
| CME | 20211104 21:11 | 20211105 11:00–23:59 | $11.4^{13.5}_{9.16}$ | 468.1 | 108.7 | ⋯ | x | ⋯ | x | Good et al. (2023) |
| CME | 20211107 18:05 | 20211109 16:10–20211110 05:30 | $7.34^{9.98}_{4.71}$ | 446.0 | 90.0 | x | x | x | x | Ledvina et al. (2023) |
| CME | 20220127 19:15 20220128 13:48 | 20220128 07:50–20220129 23:59 | $10.2^{13.6}_{6.83}$ | 364.2 | 143.1 | x | x | x | x | Ledvina et al. (2023) |
| CME* | ⋯ | 20220205 00:00–16:00 | $6.82^{8.97}_{4.67}$ | 341.7 | 123.9 | x | x | ⋯ | ⋯ | ⋯ |
| CME | 20220318 14:37 20220319 13:22 20220321 13:12 20220321 14:02 20220321 19:07 | 20220318 04:10–20220323 23:59 | $9.80^{12.0}_{7.58}$ | 276.7 | 132.6 | x | x | ⋯ | x | Mason et al. (2023); Zimbardo et al. (2023) |
| CME* | ⋯ | 20220327 03:10–09:00 | $5.88^{6.95}_{4.81}$ | 338.5 | 147.4 | x | x | ⋯ | ⋯ | ⋯ |
| CME | 20220418 23:16 | 20220418 16:30–23:30 | $4.54^{5.94}_{3.14}$ | 370.4 | 162.8 | x | ⋯ | x | x | ⋯ |

**Table 1**
(Continued)

| Category | Eruption Date (YYYYMMDD 00:00UT) | Observed at Parker (YYYYMMDD 00:00UT) | $A_{\mathrm{He}-\sigma}^{+\sigma}$ (%) | $v_{\mathrm{bulk}}$ (km s$^{-1}$) | Heliocentric Distance ($R_\odot$) | Helical Structure | Bidirectional Electrons | Helio4Cast | CCMC | Notes and References |
|---|---|---|---|---|---|---|---|---|---|---|
| CME | 20220501 14:37<br>20220502 18:43 | 20220503 06:10–<br>20220505 11:30 | $7.41_{4.58}^{10.2}$ | 429.5 | 144.8 | x | x | x | x | ⋯ |
| CME | 20220510 18:04 | 20220511 19:30–23:59 | $8.65_{6.83}^{10.5}$ | 389.2 | 127.8 | x | x | ⋯ | x | Listed SEP event: Mitchell et al. (2023) |
| CME | 20220519 16:03 | 20220520 05:00–23:59 | $6.45_{3.28}^{9.60}$ | 403.5 | 93.5 | ⋯ | x | x | ⋯ | Listed SEP event: Mitchell et al. (2023) |
| CME | 20220626 16:11<br>⋯ | 20220701 06:00–<br>20220706 09:50 | $9.01_{6.68}^{11.3}$ | 335.9 | 150.9 | x | x | x | x | ⋯ |
| CME | 20220715 16:52<br>20220716 13:46<br>20220717 14:01 | 20220718 08:00–<br>20220720 07:10 | $8.88_{6.49}^{11.3}$ | 471.4 | 163.5 | x | x | x | x | ⋯ |
| CME | 20220815 16:33<br>20220815 20:42<br>20220816 17:10<br>20220816 20:55 | 20220816 00:00–<br>20220817 11:00 | $7.65_{2.96}^{12.3}$ | 473.6 | 124.5 | x | ⋯ | ⋯ | x | ⋯ |
| CME* | ⋯ | 20220830 18:30–23:59 | $12.1_{9.28}^{14.8}$ | 517.4 | 65.2 | x | x | ⋯ | ⋯ | ⋯ |
| CME | 20220909 20:09 | 20220912 20:50–23:59 | $4.86_{2.35}^{7.38}$ | 360.3 | 62.2 | x | x | x | x | ⋯ |
| CME* | ⋯ | 20220923 00:00–06:00 | $8.11_{6.58}^{9.67}$ | 289.9 | 113.0 | x | x | ⋯ | ⋯ | ⋯ |
| CME | 20220925 16:16 | 20220927 07:10–23:59 | $23.1_{17.3}^{28.8}$ | 408.8 | 127.1 | x | ⋯ | ⋯ | x | ⋯ |
| SW | ⋯ | 20221016 00:00–09:50 | $4.62_{3.38}^{5.87}$ | 372.4 | 160.6 | x | ⋯ | ⋯ | ⋯ | Near HCS crossing |
| SW | ⋯ | 20221021 00:00–18:30 | $8.82_{6.55}^{11.1}$ | 328.7 | 163.1 | x | ⋯ | ⋯ | ⋯ | Near HCS crossing |
| CME | 20221127 20:26 | 20221129 00:00–23:59 | $5.73_{3.72}^{7.72}$ | 329.8 | 93.2 | x | x | ⋯ | x | ⋯ |
| CME | 20221203 18:18<br>20221204 10:41 | 20221205 06:00–23:59 | $5.59_{4.04}^{7.14}$ | 363.4 | 60.4 | x | ⋯ | ⋯ | x | ⋯ |
| CME* | ⋯ | 20221225 06:10–23:59 | $6.87_{4.74}^{9.02}$ | 485.5 | 101.5 | x | x | ⋯ | ⋯ | ⋯ |

**Notes.** Left to right: Categorization of either ambient Solar Wind (SW), CME, or CME*, see text for categorization scheme, if the period is associated with a CME, we Include the Eruption Date at the Sun. We list the date of the helium content observed at Parker, with the mean helium abundance with upper and lower $1\sigma$ values for the entire period. Additionally, for CMEs we note if the presence of a helical magnetic field structure is observed and/or bidirectional electrons, as well as whether the period is listed as a CME in the Helio4Cast catalog or if the CME is simulated by DONKI via CCMC, with a column for any additional notes such as publications connected to any of the events.

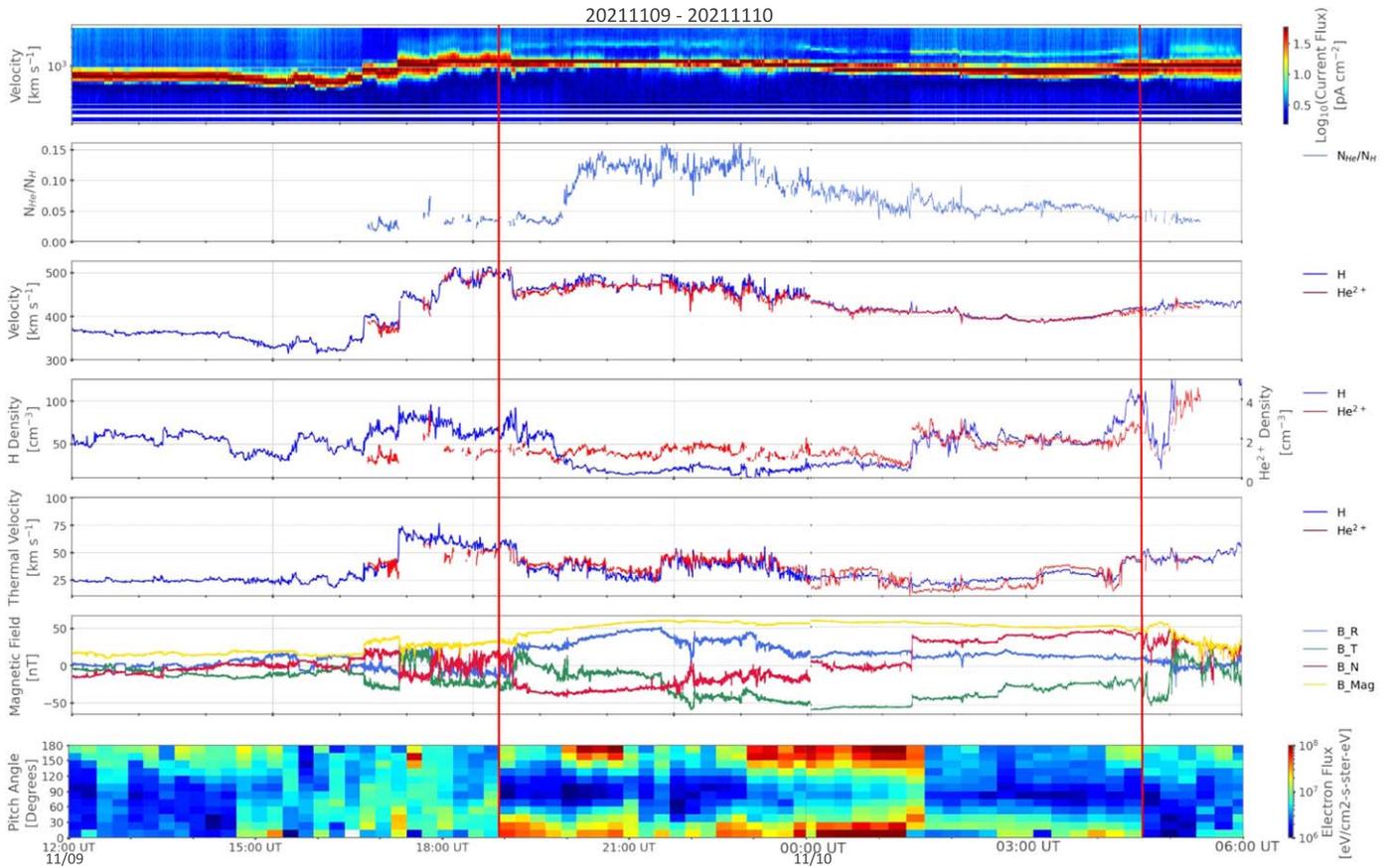

**Figure 2.** Properties of a CME from its passage on 2021 November 9 and 10 identified by the Helio4Cast catalog. Top to bottom: current, $N_{\rm He}/N_{\rm H}$, individual proton and He$^{2+}$ densities, speed, and thermal speed, magnetic field components ($B_R$, $B_T$, $B_N$, $B_{\rm Mag}$), and electron pitch-angle distribution at 314.45 eV. The vertical red lines indicate the start and end of the CME structure as identified in the Helio4Cast database.

Additionally, the pitch-angle distribution of the electrons in the bottom panel show bidirectional electrons, with enhancements at near 0° and 180°, which is a classic CME signature (Gosling et al. 1987). The counter-streaming electrons indicate a magnetic field that remains connected to the Sun at both ends.

### 4.2. Solar Wind

Occasionally, we also observe strong He$^{2+}$ signatures in ambient solar wind. Figure 3 shows an example of a persistent He$^{2+}$ signature spanning nearly a day prior to a current sheet crossing. The HCS is formed starting at the cusps of the helmet streamers across the equatorial region of the Sun and stretches throughout the entire heliosphere. The sheet of current separates the different magnetic sectors of the Sun of opposite magnetic polarity. Therefore, HCS crossings are observed in situ through changes in the local magnetic field polarity as the spacecraft crosses different hemispheres that are generally accompanied by changes in the electron strahl direction, as is observed in the bottom two panels of Figure 3.

In this period, several hours prior to the HCS crossing, we observe high helium abundances (mean value of 8.82%) associated with relatively slow speed wind ($\sim$300 km s$^{-1}$). Although no CME was identified to cross Parker during this period from the Helio4Cast catalog or DONKI simulations, the helium abundances resemble those of typical transient structures. We also observe several discontinuities in the bulk properties, including helium abundances, and field across this period, as well as a moderate rotation in the $B_R$ and $B_N$ magnetic field components that suggest the passage of a flux rope. However, given the identification as a transient would require an in-depth analysis of this individual event that is out of the scope of this work, we label this period as ambient wind as no simulation or definitive CME has been connected to this time.

We note that the signatures of elevated helium abundances appear before the HCS crossing and not explicitly during the crossing, suggesting outflowing plasma at the edge of the streamer is enriched in helium and not specifically the current sheet itself, as has been reported in other studies (Rivera et al. 2021; Lynch et al. 2023).

Figure 4 shows another example of high helium abundance in ambient wind, in this case not in the immediate vicinity of the HCS. Similarly, this event contains significantly high helium abundances not found in typical solar wind. Helium abundances are steadily 2%–3% between 09:00 and 18:00 UT, followed by a steady rise to 10% across 6 hr toward the end of the day. In connection to the He$^{2+}$ appearance in SPC, there is a gradual decrease in the velocity, density, and temperature of the protons, while the helium population increases in temperature and density during this period. We also observe no differential streaming between the protons and alpha-particle population.

Although no clear bidirectionality is observed, the electron pitch-angle distributions contain some complex structure between 06:00–12:00 UT in connection to a polarity reversal around 06:00–09:00 UT. The variability in the magnetic field and electron distributions coincide with the initial appearance of the He$^{2+}$ signature, at ∼08:00 UT, suggesting its association to the magnetic obstacle. Similar to the example in Figure 3, this case is not clearly associated with a transient, as would be indicated by the Helio4Cast catalog or DONKI simulation, therefore we log it as ambient wind.

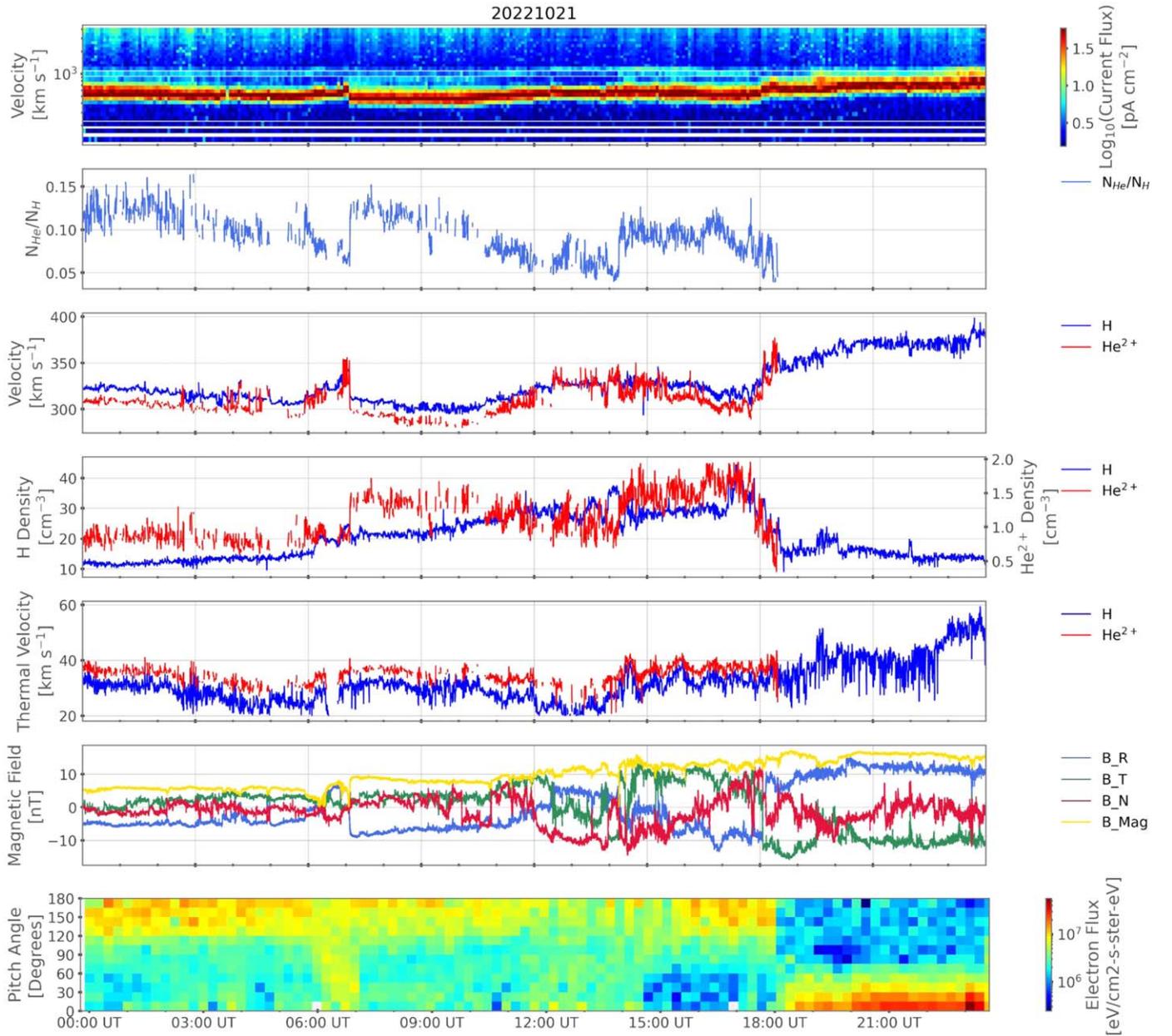

**Figure 3.** Properties of a solar-wind stream near a current sheet crossing measured on 2022 October 11; panels are the same as Figure 2.

This once again emphasizes that these initial categorizations are meant to provide some distinction between ambient and transient phenomena but may not be definitive. Many of these events show interesting features that can be further investigated.

### 4.3. Characteristic Properties of Cataloged Events

To further investigate the properties of the events listed in the table, we compare the mean helium abundance values against mean bulk speed, distance from the HCS, and across the solar cycle as indicated by the sunspot number, shown in the top-left, top-right, and bottom panels of Figure 5, respectively. The plots include the CME or CME* events listed in red and the solar wind in blue. They also include the photospheric helium abundance as a dashed black line (8.4%), and slow (1.56%) and fast speed (2.86%) mean helium abundance values during solar minimum as indicated in Song et al. (2022) as reference markers. The top-left plot includes a fitted line to the data, showing the general trend of helium abundance with mean speed. The bottom panel includes the smoothed sunspot number in green between 2018 and 2023.

When comparing the helium abundance to the mean bulk plasma velocity throughout the period, all periods together show some positive correlation, with a correlation coefficient of ∼0.4, as shown in the top-left plot of Figure 5. The helium abundance is generally more elevated for plasma traveling faster compared to that traveling at slower speeds. However, we note that the CME observed in 2002 September 25, associated with the highest $A_{\rm He} = 23.2\%$, does not indicate a markedly higher speed compared to the other events. This suggests that speed may not be a strong indicator of helium content. Given that Parker surveys the inner heliosphere, with later encounters reaching $13.3\,R_\odot$, some acceleration is expected across this region. However, we find the majority of the analyzed periods are well above this distance, the closest event being $\sim 42\,R_\odot$ with the majority being closer to $100\,R_\odot$, therefore only small acceleration effects, compared to its asymptotic speed, are expected after that point. This suggests that CMEs with the largest acceleration or highest speeds may be enriched in helium abundances compared to their less energetic counterparts.

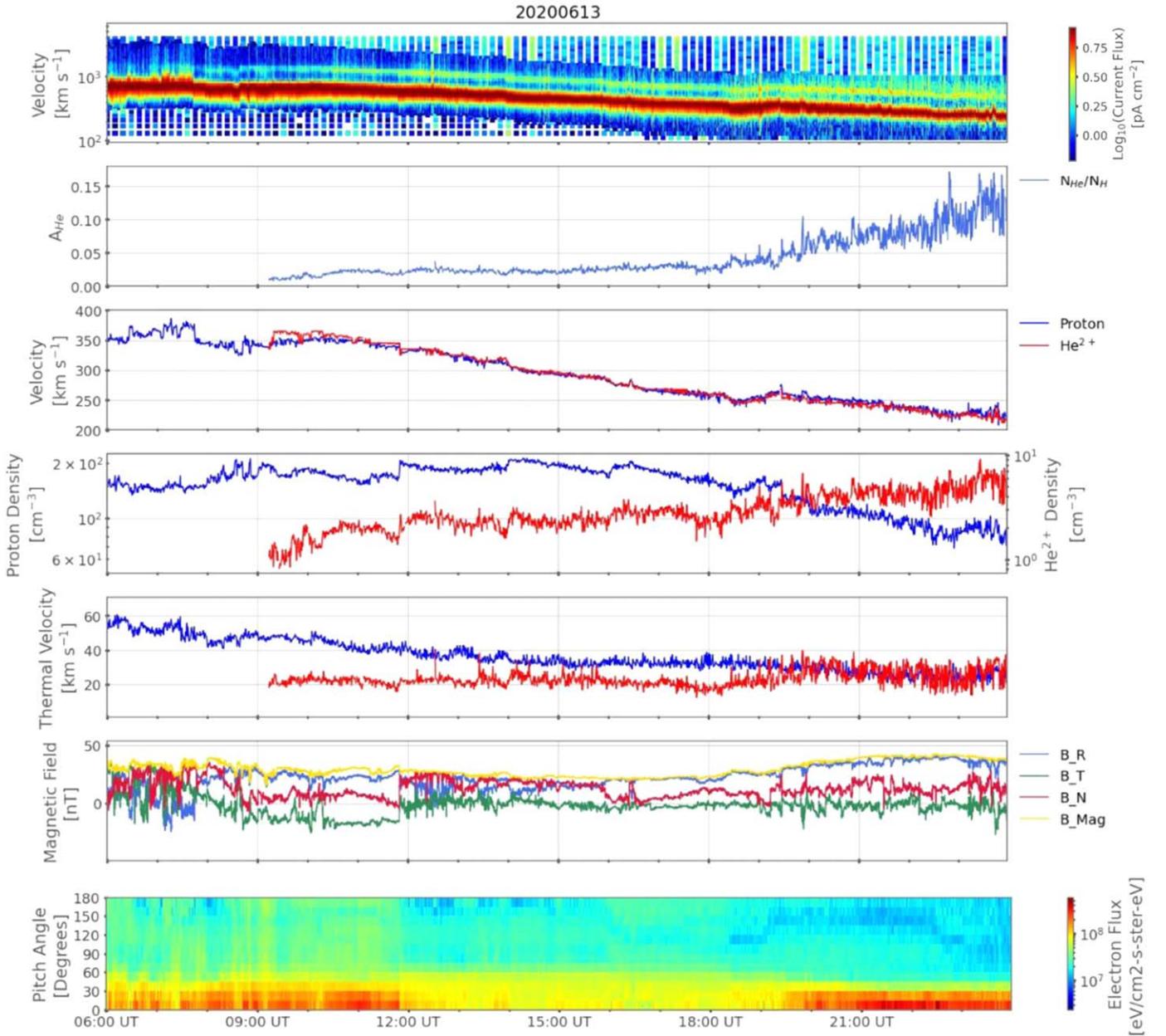

**Figure 4.** Properties of a solar-wind stream with high helium content measured on 2020 June 13; panels are the same as Figure 2.

We also investigated the angular distance from the HCS in connection to helium abundances. The HCS crossings were determined by running daily potential field source surface (Altschuler & Newkirk 1969; Schatten et al. 1969) models over the course of the Parker mission (using ADAPT/GONG magnetograms and a source surface height of $2.5\,R_\odot$) and extracting the neutral line at the outer boundary. The location of Parker relative to the neutral line was estimated using ballistic mapping at an hourly cadence over the same time interval following Badman et al. (2020), and the angular distance was estimated as described in Szabo et al. (2020).

As shown in the top-right plot of Figure 5, when examining the angular distance from the HCS to the mean helium abundance there is not a clear correlation between the location of the HCS and helium abundance values. However, we find that the frequency of events do cluster around HCS crossings, while the mean helium abundance of the events is generally lower near the HCS. Interestingly, the ambient solar-wind events are mostly clustered around the HCS, which could indicate that these events are related to streamer belt wind while the CMEs are more widespread and can blow off near or far from the HCS. These results may also indicate that CMEs are highly guided by the HCS given that they appear more frequently near it. However, given that the abundances observed are likely a subset of CMEs with the highest helium content plasma, the plot may indicate that this subset of CMEs are originating closer to the streamer belt.

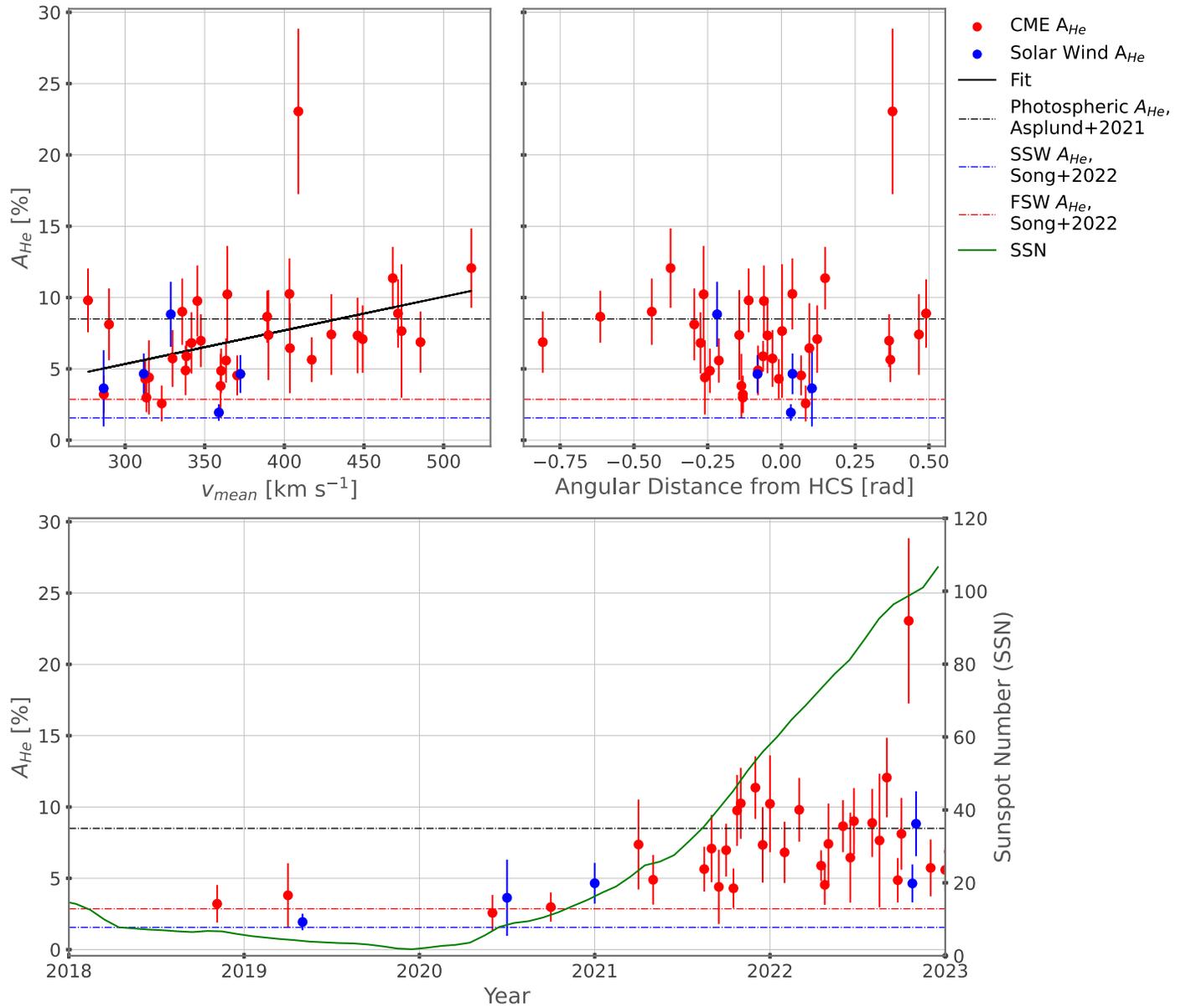

**Figure 5.** Upper left shows the relationship between transient (red) and ambient (blue) helium abundances and their bulk plasma velocity. Upper right shows the relationship between the helium abundances and angular distance from the heliospheric current sheet. Lower panel shows monthly smoothed sunspot number (SSN) in comparison with the helium abundance of helium in CMEs (in red) and other ambient solar wind (blue). All compared with mean solar maximum helium abundances of the fast solar wind (FSW) of 2.86%, slow solar wind (SSW) of 1.56%, and photospheric helium abundances of 8.4% (Asplund et al. 2021; Song et al. 2022).

Additionally, we compared the helium abundances of the listed periods in connection to the solar cycle through a comparison to the sunspot number, as shown in the bottom of Figure 5. Parker was launched at solar minimum and has been observing the ascending phase of the solar cycle. In general, there is an increase in the frequency of helium observations with increasing sunspot number in connection to an increasing number of CMEs. The higher number of observed helium observed within CMEs is expected as the Sun becomes more active. We also observe that not only do the amount of CMEs increase but also the helium abundance of some of these events increases well above the photospheric composition, shown in the horizontal dashed black line, and solar minimum numbers for the few CMEs observed. We also note that the CME with the highest mean helium abundance is observed closest to solar maximum, as would be expected by the solar-cycle dependence observed in Song et al. (2022).

Figure 6 shows the distributions of helium abundances from the CME and solar-wind periods in Table 1 with references to the photospheric, slow, and fast speed $A_{\rm He}$ values. The identified periods of the solar wind and CMEs indicate that all combined periods show an average value of $A_{\rm He} = 8.34\%$, while within identified CMEs the mean increases slightly to 8.54%, and ambient solar wind decreases to 6.90%. However, all distributions show enriched values compared to the helium abundances determined for typical slow solar wind and the fast solar wind at 1 au (Song et al. 2022). In addition, the distribution of points in the ambient wind panel (right) shows a double-peaked profile near solar-wind values, at lower $A_{\rm He}$, and at higher CME-like values, at ∼8%. The bimodal peaks are likely due to the contribution of unidentified CME plasma to this category.

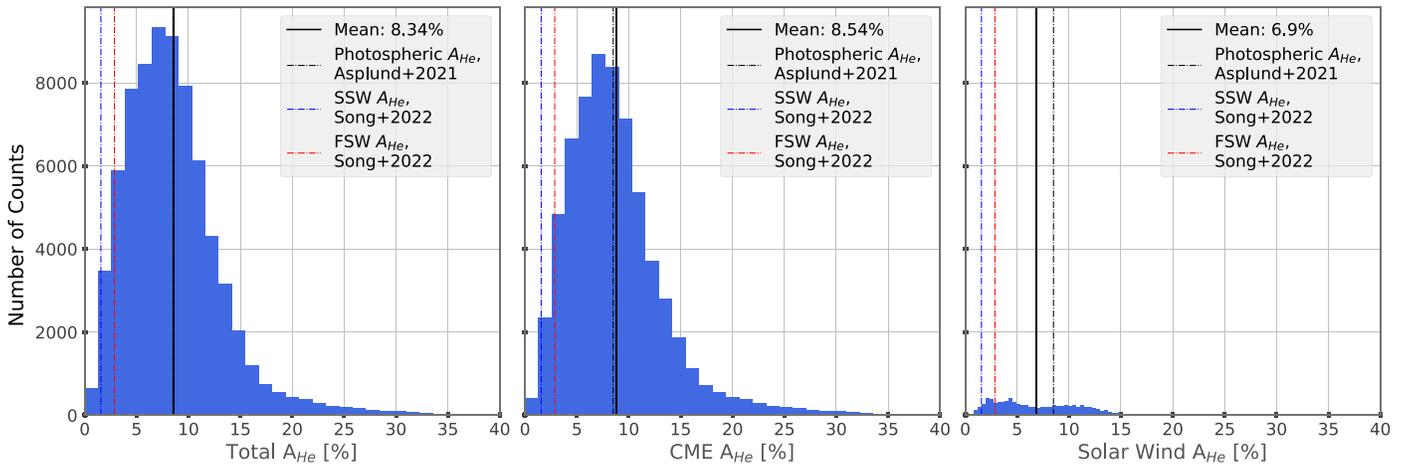

**Figure 6.** Distribution of helium abundances within (left) all events, (middle) CMEs, and (right) ambient events. The vertical lines shown in the plots are references to the helium abundances of the fast solar wind (FSW) and slow solar wind (SSW) from Song et al. (2022), 2.86% and 1.56%, respectively, and photospheric helium abundances of 8.4% from Asplund et al. (2021).

Nevertheless, the distributions are consistent with studies showing similar characteristics to the passage of CMEs versus ambient wind at 1 au that indicate higher helium content in CMEs and a general enhancement of ejecta compared to photospheric values, although the small number of solar-wind periods in our study are more elevated in helium abundances than what is typical of solar wind.

## 5. Conclusions

Through the analysis of measurements taken by SPC on Parker for encounters 1–14, this work characterizes periods of a pronounced and clearly identified population of $He^{2+}$ ions measured at the inner heliosphere. In summary, the cataloged list of events show the following:

1. Many periods show elevated mean helium abundances compared to typical solar-wind conditions of $A_{\rm He}$ values from 1%–5% (Kasper et al. 2007), as indicated in Table 1.
2. Periods identified as CMEs are the most helium-rich, with an average value of $A_{\rm He} = 8.54\%$, relative to non-CME periods (average 6.90%), as shown in Figure 6.
3. SPC is sensitive to the highest helium abundances, therefore the subset of CMEs identified are enriched compared to previous results at solar maximum and minimum, 2.5% and 6.0%, respectively, and similar to the photospheric helium abundance of 8.4% (Zurbuchen & Richardson 2006; Zurbuchen et al. 2016; Song et al. 2022).
4. The highest helium abundance observed is 23.1%, which significantly exceeds both the fast wind and slow wind, and photospheric abundances.
5. The helium content and number of both ambient and transient events increases as the solar cycle progresses into solar maximum.
6. Helium abundances show a slight dependence with increasing bulk speed.
7. Ambient solar-wind events (non-CMEs) are clustered near the HCS and could likely be associated with streamer belt outflows.

As discussed in previous studies (Richardson & Cane 2004; Zurbuchen & Richardson 2006; Richardson & Cane 2010) and found here, the plasma's high helium content is a complementary and robust manner to locate CMEs within the continuous flow of solar wind observed at Parker. Unlike other properties of CMEs that may be less developed in the inner heliosphere compared to CMEs observed in the outer heliosphere, i.e., sheath, shock structure, or expansion profile in the bulk properties, the general characteristics of elemental composition within the ejecta will remain the same at any point beyond the corona. This suggests that the elemental properties may allow for more consistent comparison of features in CMEs throughout the heliosphere. As such, we hope the comprehensive list in Table 1 provides a useful reference of interesting events for the community to investigate and explore further.

## Acknowledgments

This work is supported by the NSF-REU solar physics program at the Smithsonian Astrophysical Observatory, grant No. AGS-1850750. Partial funding for this project was supplied by the Parker Solar Probe project through the SAO/SWEAP subcontract 975569.

Parker Solar Probe was designed, built, and is now operated by the Johns Hopkins Applied Physics Laboratory as part of NASAs Living with a Star (LWS) program (contract NNN06AA01C). Support from the LWS management and technical team has played a critical role in the success of the Parker Solar Probe mission.

The FIELDS instrument suite was designed and built and is operated by a consortium of institutions including the University of California, Berkeley, University of Minnesota, University of Colorado, Boulder, NASA/GSFC, CNRS/LPC2E, University of New Hampshire, University of Maryland, UCLA, IFRU, Observatoire de Meudon, Imperial College, London, and Queen Mary University London.

The SWEAP Investigation is a multi-institution project led by the Smithsonian Astrophysical Observatory in Cambridge,

Massachusetts. Other members of the SWEAP team come from the University of Michigan, University of California, Berkeley Space Sciences Laboratory, The NASA Marshall Space Flight Center, The University of Alabama Huntsville, the Massachusetts Institute of Technology, Los Alamos National Laboratory, Draper Laboratory, JHUs Applied Physics Laboratory, and NASA Goddard Space Flight Center.

Simulation from the Space Weather Database Of Notifications, Knowledge, Information (DONKI) have been provided by the Community Coordinated Modeling Center (CCMC) at Goddard Space Flight Center through their publicly available simulation services (https://ccmc.gsfc.nasa.gov).

M.J. acknowledges and gives thanks to REU and internship mentor, Dr. Yeimy Rivera, who made this work possible. M.J. thanks the SAO/SWEAP solar wind group at the Center for Astrophysics | Harvard & Smithsonian for their incredible mentoring and guidance.

### ORCID iDs

Madisen Johnson https://orcid.org/0009-0004-0577-2496
Yeimy J. Rivera https://orcid.org/0000-0002-8748-2123
Tatiana Niembro https://orcid.org/0000-0001-6692-9187
Kristoff Paulson https://orcid.org/0000-0002-5699-090X
Samuel T. Badman https://orcid.org/0000-0002-6145-436X
Michael L. Stevens https://orcid.org/0000-0002-7728-0085
Isabella Dieguez https://orcid.org/0000-0003-4988-2967
Anthony Case https://orcid.org/0000-0002-3520-4041
Stuart D. Bale https://orcid.org/0000-0002-1989-3596
Justin Kasper https://orcid.org/0000-0002-7077-930X